**Strongly Correlated Chern Insulators in Magic-Angle Twisted Bilayer Graphene**

Kevin P. Nuckolls[1,*], Myungchul Oh[1,*], Dillon Wong[1,*], Biao Lian[2], Kenji Watanabe[3], Takashi Taniguchi[4], B. Andrei Bernevig[1], Ali Yazdani[1,‡]

[1]*Joseph Henry Laboratories and Department of Physics, Princeton University, Princeton, NJ 08544, USA*
[2]*Princeton Center for Theoretical Science, Princeton University, Princeton, NJ 08544, USA*
[3]*Research Center for Functional Materials, National Institute for Materials Science, 1-1 Namiki, Tsukuba 305-0044, Japan*
[4]*International Center for Materials Nanoarchitectonics, National Institute for Materials Science, 1-1 Namiki, Tsukuba 305-0044, Japan*

* These authors contributed equally to this work.
‡ Corresponding author email: yazdani@princeton.edu

**Interactions among electrons and the topology of their energy bands can create novel quantum phases of matter. Most topological electronic phases appear in systems with weak electron-electron interactions. The instances where topological phases emerge only as a result of strong interactions are rare, and mostly limited to those realized in the presence of intense magnetic fields[1]. The discovery of flat electronic bands with topological character in magic-angle twisted bilayer graphene (MATBG) has created a unique opportunity to search for new strongly correlated topological phases[2–9]. Here we introduce a novel local spectroscopic technique using a scanning tunneling microscope (STM) to detect a sequence of topological insulators in MATBG with Chern numbers C = ±1, ±2, ±3, which form near $\nu$ = ±3, ±2, ±1 electrons per moiré unit cell respectively, and are stabilized by the application of modest magnetic fields. One of the phases detected here (C = +1) has been previously observed when the sublattice symmetry of MATBG was intentionally broken by hexagonal boron nitride (hBN) substrates, with interactions playing a secondary role[9]. We demonstrate that strong electron-electron interactions alone can produce not only the previously observed phase, but also new and unexpected Chern insulating phases in MATBG. The full sequence of phases we observed can be understood by postulating that strong correlations favor breaking time-reversal symmetry to form Chern insulators that are stabilized by weak magnetic fields. Our findings illustrate that many-body correlations can create topological phases in moiré systems beyond those anticipated from weakly interacting models.**

The role of topology in the electronic properties of moiré flat-band systems has been experimentally revealed by the discovery of a quantized anomalous Hall conductance $\sigma_{xy} = Ce^2/h$, with various integer Chern numbers *C* in different graphene-based heterostructures[9–12].

Thus far, these topological phases have been mostly reported for device geometries that purposefully break a spatial symmetry of the graphene system. For MATBG, the alignment of graphene with its hBN substrate breaks $C_2$ symmetry, gapping the Dirac points that connect the conduction and valence flat bands of this system to form a set of degenerate flat single-particle valley Chern bands. Electron-electron interactions can favor lifting the degeneracy of these bands, resulting in spin- and/or valley-polarized insulators with Chern numbers that are consistent with those observed in experiments[9,13]. This weakly interacting picture can not only explain the reports of a C = +1 insulating phase near filling factor $\nu$ = +3 (where $\nu$ labels the number of filled moiré flat bands away from charge neutrality) observed in samples with hBN alignment, but even Chern states with C = -1 at $\nu$ = -3 and C = -2 at $\nu$ = -2 reported in samples where hBN was not suspected to be aligned with MATBG[14]. Here we report the implementation of a new experimental approach for detecting topological insulating phases and their associated Chern numbers by using density-tuned scanning tunneling spectroscopy (DT-STS) performed in the presence of a magnetic field. Our results show that strong correlations, as opposed to single-particle symmetry-breaking effects, can intrinsically produce topological phases in MATBG. Our sequence of strongly correlated Chern insulator (SCCI) phases remarkably includes not only those previously induced by hBN alignment, but also new C = ±3 phases near $\nu$ = ±1. The observed sequence of SCCI states is incompatible with the weakly interacting theories proposed thus far. However, it can be understood by postulating that interactions induce a Haldane mass term, which breaks time-reversal (T) symmetry in MATBG and modifies the single-particle picture of this system by gapping its Dirac points[15]. This mechanism produces the Chern bands necessary to explain the sequence of SCCI states we observe and can be further corroborated by our spectroscopic measurements. Our results demonstrate that strong correlations in moiré flat-band systems can produce novel correlated topological phases of matter.

     A schematic of our experiment is shown in Fig. 1a, in which we combine STM imaging of MATBG samples together with DT-STS measurements of the differential conductance d$I$/d$V$($V_s$, $V_g$) between the STM tip and the sample as a function of sample bias $V_s$ and back-gate voltage $V_g$ (Fig.1 b-e) (see methods for sample fabrication). DT-STS characterizes the local density of states (LDOS) as a function of energy and carrier concentration, which provides a powerful approach for studying phases that form in two-dimensional (2D) materials at particular carrier densities. The experiments were carried out in a homebuilt millikelvin STM and in the presence of a perpendicular magnetic field (up to 9 T)[16]. STM topographic images of the moiré superlattice confirm that our samples are near the magic angle (Fig. 1a) and are not aligned

with the hBN flake underneath (see Supplementary Section A for high-resolution images). Fig. 1b shows representative spectra measured at the center of an AA-stacked region of the moiré superlattice (bright spots in Fig. 1a)[17–19], which appear as two sharp peaks in the LDOS when the fourfold-degenerate conduction and valence flat bands of MATBG are either full (red; $\nu \approx +4$) or empty (blue; $\nu \approx -4$), respectively. At partial fillings of the flat bands (-4 < $\nu$ < +4), previous DT-STS studies have shown that the flat bands' sharp peaks in the LDOS, which identify the system's van Hove singularities (vHs), develop substantial broadening and are energetically split into a cascade of sub-band features, both of which are caused by strong electron-electron interactions[20–22]. These previous studies demonstrate that the strength of interactions in MATBG exceeds the non-interacting bandwidth of its flat bands—an important property that likely drives the formation of the topological phases we have uncovered in this study.

At millikelvin temperatures, DT-STS measurements of MATBG reveal features similar to those found at higher temperatures[20–25], but here we focus on the gaps at the Fermi level ($E_F$) detected in such measurements. Figs. 1c-e show a rich array of gaps that sequentially open and close at $E_F$ and develop with increasing magnetic field at partial flat-band fillings. The most prominent gaps are accompanied by sharp changes in the measured energies of spectroscopic features observed away from $E_F$ in the DT-STS data, signaling that the system undergoes sudden changes in its chemical potential at those fillings[21]. In the present study, we focus on gaps that occur at densities that systematically shift with the magnetic field strength. Specifically, we observe nine gaps (e.g. -15 V < $V_g$ < 10 V in Fig. 1d; marked by green triangles) about the charge neutrality point (CNP; $V_g \approx 1.5$ V), as well as six gaps away from the CNP (e.g. $V_g$ = -32.3 V, -23.6 V, -15.6 V, 18.1 V, 26.2 V, and 34.4 V in Fig. 1d; marked by red and blue triangles), all of which shift along the $V_g$ axis as we vary the field. Gaps that do not shift in density systematically as we vary the magnetic field (identified in Supplementary Section B) will be discussed elsewhere.

To examine the properties of the gaps of interest, we take a closer look at their behavior in DT-STS data measured at 6 T in Fig. 2. First, we consider the behavior of the nine gaps near the CNP (Fig. 2a), which are centered around $E_F$. Each gap is centered at a gate voltage corresponding to a carrier density, $\nu_{LL}B/\Phi_0$, where $B$ is the magnetic field, $\Phi_0 = h/e$ is the magnetic flux quantum, and $\nu_{LL}$ is an integer between -4 and 4, indicating that the gaps separate eight Landau levels (LL). These eight LLs (labeled "ZLL" in Fig. 2a) appear as peaks in *dI/dV* spectra (Fig. 2b). Since we observe two peaks when the LLs are fully filled ($V_g$ = 9.9 V) or emptied ($V_g$ = 6.6 V), we conclude that the eight LLs are grouped into two sets of four. Each fourfold-degenerate LL peak splits as it is partially filled, producing the gaps at $\nu_{LL}B/\Phi_0$.

These eight LLs can be understood as the zeroth Landau levels (ZLL)[26] that originate from the eight (spin, valley, and moiré Brillouin zone corner) Dirac cones expected in MATBG's band structure (see Supplementary Section C for spectroscopy of higher-index LLs, which are energetically well-separated from the ZLLs). This eightfold degeneracy is separated into two fourfold-degenerate manifolds by single-particle effects that could include strain[27], interlayer bias[28], and an orbital Zeeman effect[29], consistent with magnetotransport measurements that show a fourfold-degenerate Landau fan near the CNP[5–7] (see Supplementary Section D for magnetic-field dependence of ZLL peak spacing). Finally, the remaining fourfold degeneracy is spontaneously broken by exchange interactions to form quantum Hall ferromagnetic (QHF) states as the ZLL peaks are tuned to $E_F$ by the gate (Figs. 2a,b)[30,31]. Similar spectroscopic signatures of QHFs, and even real-space observations of their broken symmetry and boundary modes, have been observed in other material systems[32–34].

Next, we focus on gaps observed away from charge neutrality, two examples of which are shown in Figs. 2c-f. These gaps open and close only around $E_F$ and can therefore be identified as correlation-induced gaps (convolved with a Coulomb charging energy; Supplementary Section E). Although these gaps appear to be equally spaced in gate voltage (Figs. 1d,e), none of them occur at integer fillings $\nu$ of the flat bands at finite fields. These correlation-induced gaps, like the ZLL gaps, move along the $V_g$ axis with increasing magnetic fields, indicative of their topological nature. These gapped phases, as well as the QHF states observed around the CNP, are characterized by their Chern numbers, which we now show can be directly measured using DT-STS.

A topological gap and its associated Chern number can be identified in DT-STS measurements by studying its magnetic-field dependence. The charge density of a Chern insulating phase changes with magnetic field at a rate equal to its quantized Hall conductance[35,36], $dn/dB = \sigma_{xy}/e = C/\Phi_0$. Therefore, tuning both the carrier density and the magnetic field in DT-STS measurements remarkably allows us to identify topological gaps and their associated Chern numbers using this local spectroscopic technique. Fig. 3a shows that the gaps display this expected linear dependence in a magnetic field, all measured within Device A (see Supplementary Section F and G for Devices B and C). In Fig. 3b, we aggregate the results of such measurements from three different devices, all of which show very similar behavior. This information has been reparametrized as a function of flat-band filling $\nu$ and magnetic flux quanta per superlattice unit cell $\Phi/\Phi_0$ (see Supplementary Sections H and I for details). The error bars in these figures represent the entire density range over which each gap is observed, with markers placed at the center of each range (see Supplementary Section J for methods). The shaded regions (Fig. 3a) or lines (Fig. 3b) with different slopes identify the possible quantized Chern numbers for each gap. This figure illustrates the power of applying this straightforward analysis to high-quality DT-STS data, which we use to uncover the rich array of correlation-driven topological phases that can form in MATBG at partial fillings of its flat bands.

The sequence of Chern gaps in our experiments (Fig. 3b) includes a hierarchy of insulating phases with Chern numbers C = ±1, ±2, ±3 emanating as a function of magnetic field from filling factors $\nu$ = ±3, ±2, ±1, respectively, a set of QHF phases from the ZLLs, and a few higher-index LLs observed near the CNP. All Chern insulating phases appear to be stabilized by a magnetic field[7,14], where the C = +2 insulator is seen to develop at fields as low as 1 T ($\sim 0.04\Phi_0$), the C = ±1 around 3 T ($\sim 0.1\Phi_0$), and the C = ±3 and C = -2 around 6 T ($\sim 0.2\Phi_0$).

To understand the mechanism driving the formation of Chern insulating phases in MATBG, we start with the single-particle, low-energy electronic structure of the MATBG flat bands, schematically shown in Fig. 4a. In the non-interacting limit, MATBG has a composite $C_2T$ symmetry, which protects the Dirac points between the valence and conduction flat bands. Interactions can gap the Dirac points through a $C_2$- or T-symmetry-breaking (or both) mass term, producing degenerate, flat Chern bands with Chern number ±1. This lowers the total free energy F ≡ E – μN of the system by decreasing (increasing) the energy of occupied (unoccupied) sub-bands. Previous studies discovered a Chern insulating phase near $\nu$ = +3 with C = +1 in devices, where a $C_2$-symmetry-breaking mass was imposed externally by alignment with the hBN substrate[8,9]. This gives rise to isolated flat sub-bands in the *K* and *K'* valleys of the original graphene with opposite Chern numbers (Fig. 4b)[9]. For these samples, interactions can spontaneously break time-reversal symmetry and valley-polarize this system, creating the equivalent of a quantum Hall ferromagnet, thereby creating a single unoccupied Chern sub-band consistent with the experimentally observed C = +1 state. At first glance, our observation of a C = +1 state near $\nu$ = +3 in the absence of hBN alignment may suggest that strong interactions spontaneously generate a $C_2$-symmetry-breaking mass, giving rise to a band topology similar to that of the hBN-aligned samples. However, this mechanism is expected to produce C = ±1 states near $\nu$ = ±1 because time-reversed partner sub-bands in the *K* and *K'* valleys would have opposite Chern numbers (Fig. 4b). Such states are inconsistent with the C = ±3 states we have observed near these fillings (Supplementary Section K).

Instead of a $C_2$-symmetry-breaking mass, we alternatively postulate that the Dirac points are gapped by interactions in the form of a time-reversal-symmetry-breaking mean-field mass term. Since the Chern number decrements by 1 between each insulating gap as $\nu$ is increased, our measurements indicate repeated occupancy of bands with Chern number -1. A Haldane mass term provides exactly this by ensuring that $C_2$-related partner sub-bands in the $K$ and $K'$ valleys have the same Chern numbers (Fig. 4b)[15]. The sign of the Haldane mass determines the sign of the Chern number for each SCCI, so the Haldane mass must change sign as the system evolves from n-doped ($\nu > 0$) to p-doped ($\nu < 0$), with the mass vanishing at the CNP (Fig. 4d). In this picture, the sequential doping of the conduction and valence sub-bands gives rise to the series of experimentally observed states with Chern numbers C = ±1, ±2, ±3 near filling factors $\nu$ = ±3, ±2, ±1. We argue that a small magnetic field favors Chern phases with these extremal Chern numbers (i.e. those formed by a T-symmetry-breaking mass) and with the same sign as the product $\nu$B simply by free-energy considerations (Supplementary Section L).

We speculate two possible scenarios for the behavior of MATBG near zero magnetic field and at integer fillings. One possibility is that competing topologically trivial states are favored near zero field[4,6,7]. A second, more interesting possibility is that the system may form domains with different Chern numbers, which masquerades as a topologically trivial insulator near zero field, while the application of a weak magnetic field may favor unifying the sample into a single high-Chern number insulating domain.

In addition to producing the correct sequence of SCCI states, our picture of an interaction-induced Haldane mass that changes sign across the CNP can be further corroborated by examining the spectroscopic behavior of the ZLLs in MATBG. When a Dirac cone is gapped to form a pair of Chern bands, its corresponding ZLL is replaced by an unpaired LL that emerges from the edge of the band with Chern number +1 (for simplicity, we still refer to this LL as a ZLL)[15]. A $C_2$-breaking mass in MATBG would create conduction (valence) sub-bands with opposite Chern numbers in the K and K' valleys, thus creating a symmetric spectrum of ZLLs emerging from the conduction bands of one valley and valence bands of the other valley. In contrast, the Haldane mass yields conduction (valence) sub-bands with the same Chern number in the two valleys, thus creating an asymmetric spectrum of ZLLs that all emerge from either the valence or the conduction bands, depending on the sign of the Haldane mass (Fig. 4c). Examining the data in Fig. 2a, we find that the ZLLs are closer in energy to the valence flat band vHs when MATBG is n-doped (e.g. $V_g$ = 11 V) but are closer in energy to the conduction flat band vHs when MATBG is p-doped (e.g. $V_g$ = -7 V). This behavior is consistent with the change of the sign of the Haldane mass proposed to produce the observed sequence of Chern states in our experiments. At $V_g$ = 1.5 V, the ZLLs appear symmetric (with the single-particle splitting discussed previously) because the Haldane mass is zero at the CNP. A more detailed model, which includes localization due to both the magnetic field and the enhanced LDOS of the flat bands at the AA sites, can produce the fine features of the DT-STS data (Supplementary Section M). Finally, corroborating evidence for the Chern states we have uncovered here can also be found in early magneto-capacitance measurements of MATBG, which showed signatures of compressibility changes at some of the fillings examined here[37]. However, in absence of spectroscopic information, the previous study could not identify the origin of these features as the correlated gaps induced at $E_F$ (Figs. 2c-f) due to SCCI phases forming near these fillings.

Looking ahead, we anticipate that the strong interactions responsible for the formation of SCCI states near integer fillings of the flat bands in our MATBG devices without hBN alignment may also be capable of stabilizing fractional Chern insulators, similar to those recently proposed for aligned samples[38–40]. Furthermore, the capability of DT-STS to clearly distinguish trivial and topological gaps by measuring their Chern numbers, as demonstrated here, can be extended to study different topological phases that may form in a wide range of different 2D materials and moiré platforms[10,41–43]. A key advantage of this technique is the ability to circumvent the need for micron-scale device homogeneity or complex device geometries due to its versatility as a local probe method.

**Acknowledgements**

We thank S. Wu, B. Jaeck, X. Liu, K. Hejazi, N. Yuan, and L. Fu for useful discussions. This work was primarily supported by the Gordon and Betty Moore Foundation's EPiQS initiative grants GBMF4530, GBMF9469, and DOE-BES grant DE-FG02-07ER46419 to A.Y. Other support for the experimental work was provided by NSF-MRSEC through the Princeton Center for Complex Materials NSF-DMR-1420541, NSF-DMR-1904442, ExxonMobil through the Andlinger Center for Energy and the Environment at Princeton, and the Princeton Catalysis Initiative. K.W. and T.T. acknowledge support from the Elemental Strategy Initiative conducted by the MEXT, Japan, grant JPMXP0112101001, JSPS KAKENHI grant JP20H00354, and the CREST (JPMJCR15F3), JST. B.L. acknowledges support from the Princeton Center for Theoretical Science at Princeton University. B.A.B. acknowledges support from the Department of Energy DE-SC0016239, Simons Investigator Award, the Packard Foundation, the Schmidt Fund for Innovative Research, NSF EAGER grant DMR-1643312, NSF-MRSEC DMR1420541, BSF Israel US foundation No. 2018226, ONR No. N00014-20-1-2303, and the Princeton Global Network Funds.

**Author Contributions**

K.P.N., M.O., D.W., and A.Y. designed the experiment. K.P.N., D.W., and M.O. fabricated samples, carried out STM/STS measurements, and performed the data analysis. B.L. and B.A.B. performed the theoretical calculations. K.W. and T.T. synthesized the hBN crystals. All authors discussed the results and contributed to the writing of the manuscript.

**Figure Captions:**

**Figure 1 | Magnetic-field-dependent spectroscopic gaps in MATBG at 200 mK. a,** Schematic diagram of the experimental setup. Inset: STM topographic image of magic-angle region ($V_s$ = -98 mV, $I$ = 200 pA; Scale bar = 5 nm). **b,** d$I$/d$V$($V_s$) spectra measured at the center of an AA site at zero magnetic field. The red, blue, and purple spectra were obtained when the chemical potential was tuned to lie above, below, and between the two flat bands of MATBG, respectively. (Initial tunneling parameters: $V_s$ = -100 mV; $I$ = 500 pA; 4.121 kHz sinusoidal modulation of $V_{rms}$ = 1 mV). **c,** d$I$/d$V$($V_s$, $V_g$) measured at the center of an AA site at $B_\perp$ = 1 T. **d,**

d$I$/d$V$($V_s$, $V_g$) measured at the center of an AA site at $B_\perp$ = 6 T. **e,** d$I$/d$V$($V_s$, $V_g$) measured at the center of an AA site at $B_\perp$ = 9 T. (Initial tunneling parameters: $V_s$ = -80 mV (**c**, **d**), -70 mV (**e**); $I$ = 1.5 nA (**c**, **e**), 1 nA (**d**); 4.121 kHz sinusoidal modulation of $V_{rms}$ = 0.2 mV)

**Figure 2 | Spectroscopic gap morphology of strongly correlated Chern insulating gaps and zeroth Landau Levels. a,** d$I$/d$V$($V_s$, $V_g$) measured at the center of an AA site at $B_\perp$ = 6 T showing the ZLL crossing $E_F$ (same data as Fig. 1d). **b,** d$I$/d$V$ line cuts of data in **a** for $V_g$ = 9.9 V, 9.0 V, 7.2 V, 5.4V, 2.0V, 0.2 V, -2.3 V, -4.1V, -6.6 V (Vertical Offset = 20 - 40 nS). Highlighted numbers correspond to the degeneracy of each zeroth Landau level peak. **c,** d$I$/d$V$($V_s$, $V_g$) measured at the center of an AA site at $B_\perp$ = 6 T in a saturated color scale used to highlight a field-stabilized spectroscopic gap at $E_F$ between $\nu$ = 2 and $\nu$ = 1. **d,** d$I$/d$V$ line cuts of data in **c** from $V_g$ = 20.6 V to 13.6 V (Vertical Offset = 32 nS). Triangles mark the edges of the spectroscopic gap. **e,** d$I$/d$V$($V_s$, $V_g$) measured at the center of an AA site at $B_\perp$ = 6 T in a saturated color scale used to highlight a field-stabilized gap between $\nu$ = -1 and $\nu$ = -2. **f,** d$I$/d$V$ line cuts of data in **e** from $V_g$ = -13 V to -18.6 V (Vertical Offset = 32 nS).

**Figure 3 | Quantized magnetic-field response of strongly correlated Chern insulating phases. a,** Scatter plot of the gate voltage and magnetic field for extracted spectroscopic gaps from d$I$/d$V$($V_s$, $V_g$) measurements on one device (Device A). Purple shaded bars depict expected quantized field response of the Landau level gaps with Landau level filling factors $\nu_{LL} \in [-4, 4]$. Red and blue shaded bars depict expected quantized field response of Chern insulating gaps with C = ±1, ±2, ±3 emanating as a function of magnetic field from integer flat-band fillings $\nu$ = ±3, ±2, ±1, respectively. The width of the shaded bars is derived from the error in determining band full ($\nu$ = 4) and empty ($\nu$ = -4) fillings, between which all integer fillings $\nu$ were defined to be equally spaced. **b,** Magnetic-field response of Chern and Landau level gaps for three devices (Devices A, B, and C), reparametrized by the number of magnetic flux quanta per superlattice unit cell $\Phi/\Phi_0$ and flat-band filling $\nu$. The three solid lines emanating from each nonzero integer $\nu$ depict the expected quantized field response of insulators with Chern numbers C = ±1, ±2, ±3. Solid lines colored red or blue indicate the only Chern numbers consistent with our data.

**Figure 4 | Theoretical interpretation using an interaction-induced, sign-switching Haldane mass. a,** Schematic depiction of a non-interacting model of the band structure of MATBG. Depicted are the eightfold-degenerate Dirac points attributed to two spins, two valleys of the graphene Brillouin zone, and two corners of the moiré Brillouin zone. **b,** Schematic depiction of

the two types of broken symmetries that cause the Dirac points to gap. $C_2$ symmetry is broken by a staggered potential mass (of the form $M_S \sigma_z \otimes 1$, where $M_S$ is the mass magnitude and $\sigma_z$ is a Pauli matrix acting on the sublattice pseudospin), which opens a mass gap of the same magnitude in each valley. Time-reversal symmetry is broken by an interaction-induced Haldane mass (of the form $M_H \sigma_z \otimes \tau_z$, where $M_H$ is the Haldane mass magnitude and $\tau_z$ acts on the valley degree of freedom), which opens a mass gap of equal and opposite magnitude in the K and K' valleys. **c,** Depiction of the zeroth Landau level asymmetry observed in d*I*/d*V*(*V*$_s$), which reveal the sign-switching behavior of the Haldane mass across the CNP. **d,** Schematic depiction of the doping-dependent behavior of the Haldane mass, which must change sign across charge neutrality. Sequential filling of the Chern -1 bands (blue) and Chern +1 bands (red) up to chemical potential μ lead to the experimentally observed topological phases.

**Data Availability**

The data that supports the findings of this study are available from the corresponding author upon reasonable request.

**Competing Interests**

The authors declare no competing interests.

**Methods**

**STM measurements.** Measurements were performed on a homebuilt, ultra-high vacuum (UHV), dilution-fridge STM with base electron temperature $T_{electron} \approx 200$ mK. All tungsten tips were prepared on a Cu(111) single crystal, and subsequently calibrated against the Cu(111) Shockley surface state. MATBG devices were electrostatically gated via a degenerately p-doped Si back-gate. The sample was biased by a voltage $V_s$ relative to a virtually grounded tip. Density-tuned scanning tunneling spectroscopy (DT-STS) measurements were performed using a standard lock-in technique where the ac response to a small sinusoidal voltage $V_{rms}$ applied to the sample is measured while tuning the sample bias and gate voltage.

**Sample preparation.** Devices were prepared using a method nearly identical to the one described in Ref. 21. Briefly, MATBG devices were assembled using the "tear-and-stack" method, where a PVA/scotch-tape/PDMS/glass-slide transfer handle was first used to pick up hBN from an $SiO_2$/Si wafer. Next, hBN was used to contact half of a monolayer flake of graphene (exfoliated on a piranha-cleaned $SiO_2$/Si wafer using a hot cleave method). Graphene was torn at the hBN flake's edge and the second half of the graphene was picked up using the handle-adhered stack after a 1.2° - 1.3° relative rotation of the transfer station stage (which accounts for an observed rotation relaxation). Finally, the stack was transferred onto a PDMS/glass-slide transfer handle, which inverts the stack's vertical orientation for STM measurements, and transferred onto a prepatterned $SiO_2$/Si wafer with Au/Ti electrodes. The sample is then cleaned with water and/or various solvents before annealing in UHV for 12 hours at 170 °C, followed by a 2-hour anneal at 400 °C.

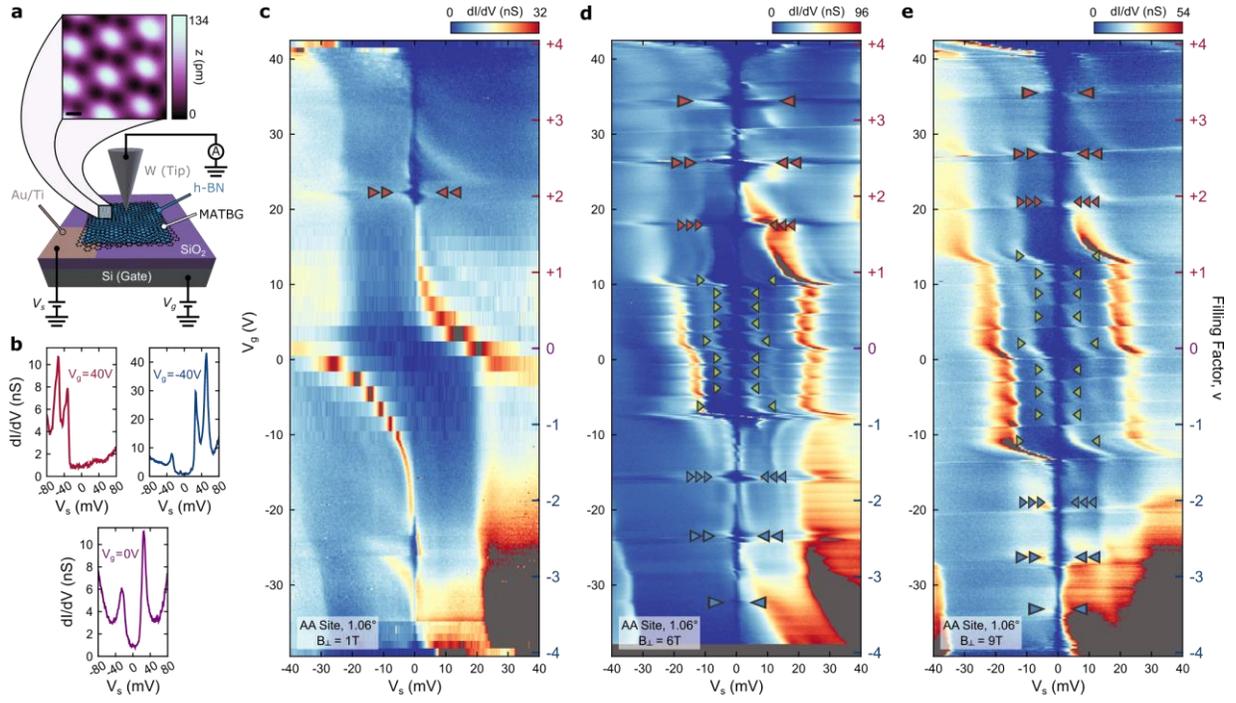

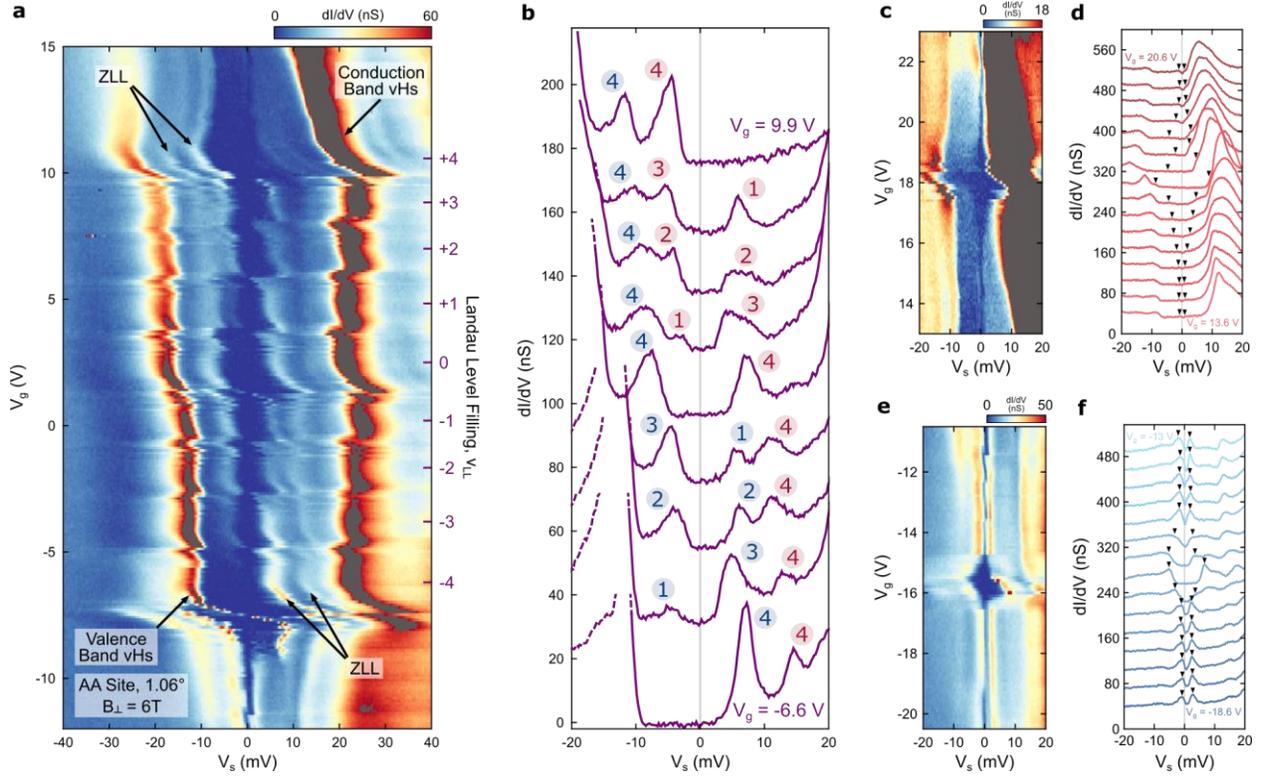

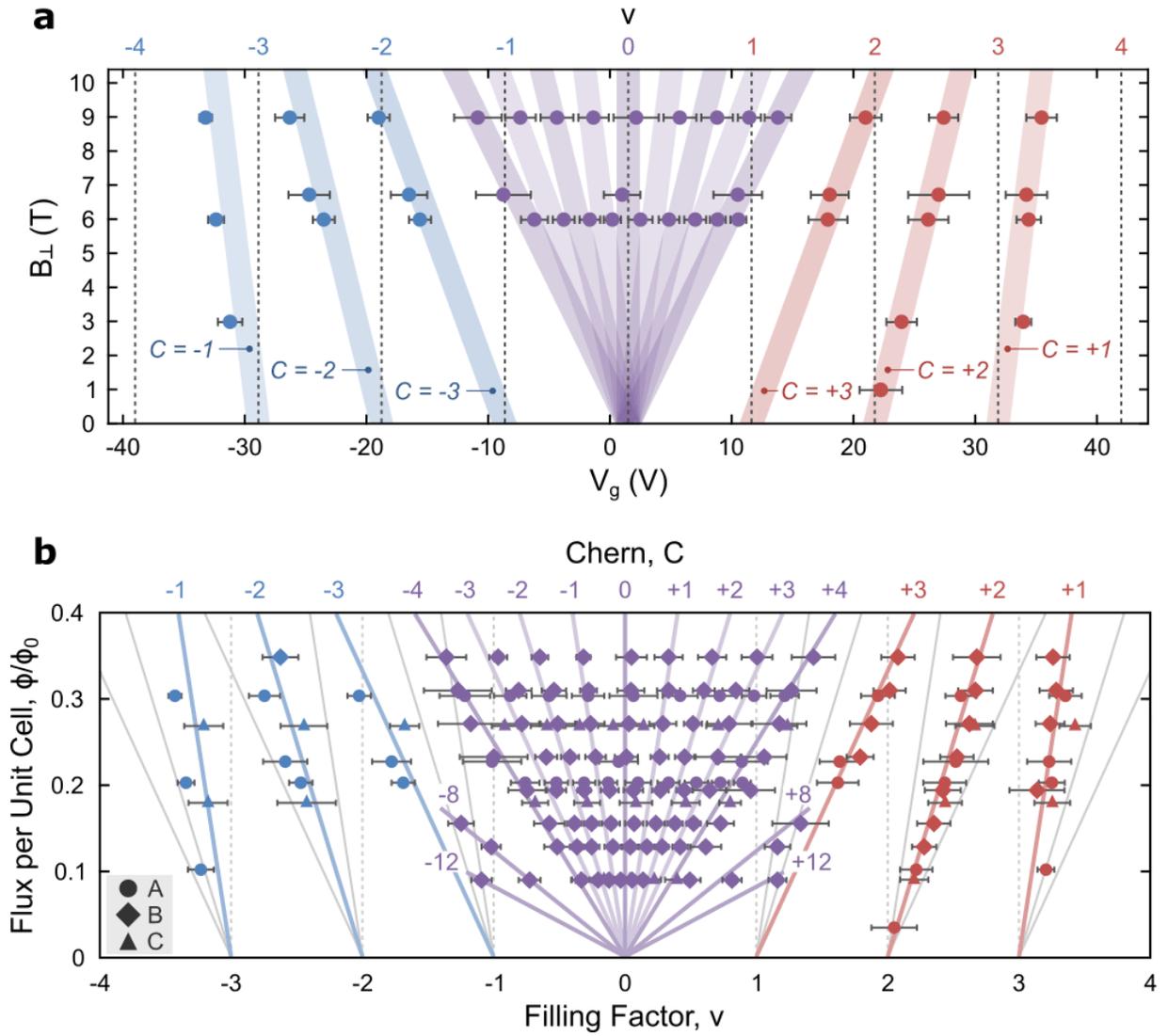

# FIGURE 4

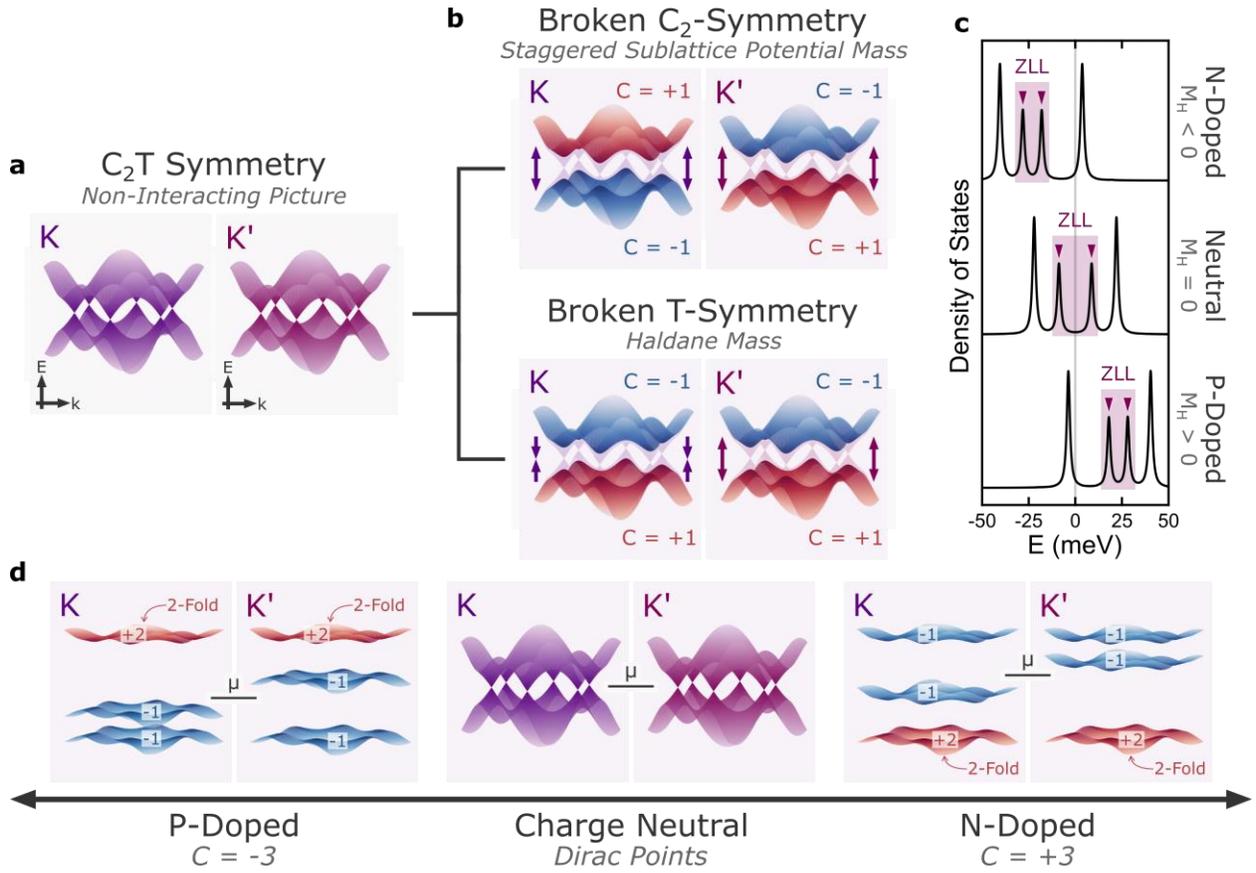